\documentclass[12pt]{amsart}
\usepackage{amssymb,latexsym}
\newdimen\AAdi%
\newbox\AAbo%
\def\AAk#1#2{\s_etbox\AAbo=\hbox{#2}\AAdi=\wd\AAbo\kern#1\AAdi{}}%
\def\AAr#1#2#3{\s_etbox\AAbo=\hbox{#2}\AAdi=\ht\AAbo\raise#1\AAdi\hbox{#3}}%
\font\tenmsb=msbm10 at 12pt \font\sevenmsb=msbm7 at 8pt
\font\fivemsb=msbm5 at 6pt
\newfam\msbfam
\textfont\msbfam=\tenmsb \scriptfont\msbfam=\sevenmsb
\scriptscriptfont\msbfam=\fivemsb
\def\Bbb#1{{\tenmsb\fam\msbfam#1}}

\usepackage{amssymb}
\begin{document}

\newcommand{\D}{D \hskip -2.8mm \slash}
\newcommand{\p}{\partial \hskip -2.2mm \slash}
\newcommand{\pp}{\overline{\psi}}
\newcommand{\e}{\overline{\varepsilon}}
\newcommand{\hh}{\sqrt{h}d^2x}
\newcommand{\ii}{\frac{1}{2}\int_M}
\renewcommand{\baselinestretch}{2}
\renewcommand{\theequation}{\thesection.\arabic{equation}}
\newcommand{\wb}{\widetilde{\nabla}_{e_\beta}}
\newcommand{\wa}{\widetilde{\nabla}_{e_\alpha}}

\newtheorem{thm}{Theorem}
\newtheorem{lem}{Lemma}
\newtheorem{cor}{Corollary}
\newtheorem{rem}{Remark}
\newtheorem{pro}{Proposition}
\newtheorem{defi}{Definition}
\newcommand{\noi}{\noindent}
\newcommand{\dis}{\displaystyle}
\newcommand{\mint}{-\!\!\!\!\!\!\int}
\newcommand{\ba}{\begin{array}}
\newcommand{\ea}{\end{array}}
\def \tf{\tilde{f}}
\def\cqfd{%
\mbox{ }%
\nolinebreak%
\hfill%
\rule{2mm} {2mm}%
\medbreak%
\par%
}
\def \pr {\noindent {\it Proof.} }
\def \rmk {\noindent {\it Remark} }
\def \esp {\hspace{4mm}}
\def \dsp {\hspace{2mm}}
\def \ssp {\hspace{1mm}}
\def\n{\nabla}
\def\C{\Bbb C}
\def\B{\Bbb B}
\def\N{\Bbb N}
\def\Q{\Bbb Q}
\def\Z{\Bbb Z}
\def\EE{\Bbb E}
\def\H{\mathbb{H}}
\def\S{\mathbb{S}}
\def \c {{\bf C}}
\def \R {\mathbb{R}}
\def \Z {{\bf Z}}
\def \Q {{\bf Q}}
\def \a {\alpha}
\def \b {\beta}
\def \d {\delta}
\def \e {\epsilon}
\def \G {\Gamma}
\def \g {\gamma}
\def \l {\lambda}
\def \L {\Lambda}
\def \O {\Omega}
\def \om {\omega}
\def \o{\omega}
\def \s {\sigma}
\def \t {\theta}
\def \z {\zeta}
\def \vp {\varphi}
\def \vt {\vartheta}
\def \ve {\varepsilon}
\def \i {\infty}
\def \ds {\displaystyle}
\def \oo {\overline{\Omega}}
\def \ov {\overline}
\def \bd {\bigtriangledown}
\def \U {\bigcup}
\def \un {\underline}
\def \h {\hspace{.5cm}}
\def \hs {\hspace{2.5cm}}
\def \v {\vspace{.5cm}}
\def \mi {M_{i}}
\def \ra {\longrightarrow}
\def \Ra {\Longrightarrow}
\def \rw {\rightarrow}
\def \bs {\backslash}
\def \rn {{\bf R}^n}
\def \h* {\hspace*{1cm}}
\def\la{\langle}
\def\ra{\rangle}
\def\cal{\mathcal}

\title[Liouville Theorems  for  Dirac-Harmonic Maps]{Liouville Theorems  for  Dirac-Harmonic Maps }

\author{
Q. Chen, J. Jost and G. Wang
}

\thanks{The research of QC is partially supported by  NSFC (Grant No.10571068) and SRF for
ROCS, SEM, he also thanks the Max Planck Institute for Mathematics
in the Sciences for good working conditions during his visit.}

\address{School of Mathematics and Statistics\\ Wuhan University\\Wuhan 430072, China } \email{qunchen@whu.edu.cn}

\address{Max Planck Institute for Mathematics in the Sciences\\Inselstr. 22\\D-04103 Leipzig, Germany} \email{jjost@@mis.mpg.de}

\address{Faculty of Mathematics, University Magdeburg, D-39016, Magdebrug, Germany}
\email{gwang@math.uni-magdeburg.de}


\begin{abstract}
We prove Liouville theorems for  Dirac-harmonic maps  from the Euclidean space $\R^n$, the hyperbolic space $\H^n$ and a Riemannian manifold $\mathfrak{S^n}$   ($n\geq 3$) with the Schwarzschild metric to any Riemannian manifold $N$.
\end{abstract}
\date{April 2, 2007}
\maketitle
{\small
 } \vskip24pt

\section{Introduction}

Let $(M^n,g)$ be a Riemannian manifold with fixed spin structure,
 $\Sigma M$ its spinor bundle, on which we chose a Hermitian
metric $\la\cdot,\cdot\ra$. Let $\n$ be the Levi-Civita connection
on $\Sigma M$ compatible with $\la\cdot,\cdot\ra$ and $g$. Let
$\phi$ be a smooth map from $M$ to a Riemannian manifold $(N,h)$ of
dimension $n'\ge 2$ and $\phi^{-1} TN$  the pull-back bundle of $TN$
by $\phi$. On the twisted bundle $\Sigma M\otimes \phi^{-1} TN$
there is a metric (still denoted by $\la \cdot,\cdot\ra$) induced
from the metrics on $\Sigma M$ and $\phi^{-1} TN$. There is also a
natural connection $\widetilde{\n}$ on $\Sigma M\otimes \phi^{-1}
TN$ induced from those on $\Sigma M$ and $\phi^{-1} TN$. In local
coordinates $\{x_\alpha\}$ and $\{y^i\}$ on $M$ and $N$
respectively, we write the section $\psi$ of $\Sigma M\otimes
\phi^{-1} TN$ as
\[\psi(x)= \psi^j (x)\otimes\partial_{y^j}(\phi(x)),\]
where $\psi^i$ is a spinor on $M$ and $\{\partial_{y^j}\}$
is the natural local basis on $N$, and  $\widetilde{\n}$ can be
written as
\[\widetilde{\nabla} \psi(x)= \n \psi^i(x)\otimes \partial_{y^i}(\phi(x))
+ \Gamma^i_{jk} \n \phi^j(x) \psi^k(x)\otimes\partial_{y^i}
(\phi(x)).\] Here and in the sequel, we use the summation
convention.
\par
The {\it Dirac operator along the map $\phi$} is defined as
 \begin{eqnarray*}
\D\psi&:=& e_\a\cdot \widetilde{\nabla}_{e_\a}
\psi\nonumber \\
&=&
\partial \hskip -2.2mm \slash
 \psi^i(x)\otimes\partial_{y^i}(\phi(x))
+ \Gamma^i_{jk} \n_{e_\a} \phi^j(x) e_\a\cdot
\psi^k(x)\otimes\partial_{y^i}(\phi(x)),
\end{eqnarray*}
 where
$\{e_\alpha\}$ is the local orthonormal basis of $M$ and $
\partial \hskip -2.2mm \slash
:=e_{\a}\cdot \n_{e_\a}$ is the usual Dirac operator on $M$.
 The Dirac
operator $\D$ is formally self-adjoint, i.e.,
\begin{equation}
\label{1.2}
 \int_M\la\psi, \D\hskip0.8mm\xi\ra=\int_M\la\D\hskip0.8mm\psi,\xi\ra,
 \end{equation}
for all $\psi, \xi \in \Gamma (\Sigma M\otimes \phi^{-1}TN)$. For
properties of the spin bundle $\Sigma M$ and the Dirac operator $
\partial \hskip -2.2mm \slash$, we refer
the readers to \cite{LM} or \cite{J}.

 Let us consider the functional
\begin{equation}
\label{b0}
L(\phi,\psi):=\frac 1 2 \int_M[|d\phi|^2+\la\psi,\D\psi\ra],
\end{equation}
where $\la\psi,\xi\ra:=h_{ij}(\phi)\la\psi^i,\xi^j\ra$, for $\psi,\xi\in
\Gamma (\Sigma M\otimes \phi^{-1}TN)$.

The Euler-Lagrange equations of $L$ are (see \cite{CJLW1}):
\begin{equation}
\label{b2}
\tau^i(\phi)=\frac{1}{2}R^i\hskip0.0003mm_{jkl}(\phi)\la\psi^k,\nabla\phi^j\cdot\psi^l\ra,
\end{equation}
\begin{equation}
\label{b1} \D\psi^i:=\partial \hskip -2.2mm \slash
\psi^i+\Gamma^i_{jk}(\phi)\partial_\alpha\phi^je_\alpha\cdot\psi^k=0,
\end{equation}
$i=1,2,\cdots,n':={\rm dim}N,$
where $\tau(\phi)$ is the tension field of the map $\phi$.
\par
Denoting
$$\cal{R}(\phi,\psi):=\frac{1}{2}R^i\hskip0.0003mm_{jkl}(\phi)\la\psi^k,\nabla\phi^j\cdot
\psi^l\ra\otimes\partial_{y^i},$$ then (\ref{b2}) and (\ref{b1}) can
be written as:
\begin{equation}
\label{b2.1} \tau(\phi)=\cal{R}(\phi,\psi),
\end{equation}
\begin{equation}
\label{b1.1} \D\psi=0.
\end{equation}
\par
We call solutions $(\phi,\psi)$ of the coupled system (\ref{b2}) and (\ref{b1})  Dirac-harmonic maps
  from $M$ into $N$. The system (\ref{b2}, \ref{b1})  arises from the supersymmetric
  nonlinear sigma model of quantum field theory by making all
  variables commuting (see \cite{CJLW1} and
  \cite{CJLW2}). Thus, Dirac-harmonic constitute a natural extension of
  the harmonic maps thoroughly studied in geometric analysis. An
  obvious question then is to what extent the structural theory of
  harmonic maps generalizes to Dirac-harmonic maps.
\par
In the present paper, our starting point in this direction is
\cite{GRSB}, where, motivated again by considerations from quantum
field theory, it was proved that any harmonic map of finite energy from the Euclidean space $\mathbb{R}^n$
($n\geq 3$) into a Riemannian manifold $N$ must be constant. In \cite{Se},
 this vanishing property was shown for the case of the domain manifold $\mathbb{H} ^n$, the hyperbolic space.
 These Liouville theorems are a consequence of the non-invariance of the energy
 functional under conformal transformations, and the fact that there exist conformal vector fields on
 the domains. In \cite {Pk}, these results
  were extended to the case where the domain is a Riemannian manifold $\mathfrak{S}^n$ with the Schwarzschild
  metric (see definitions and notations in section 3).

In contrast to harmonic maps, the integrands in the functional $L$
for Dirac-harmonic maps are not nonnegative in general, and the
energy functional should be chosen as follows (c.f. \cite{CJLW1} and
\cite{CJLW2}):
$$E(\phi,\psi):=\int_M[|d\phi|^2+|\psi|^4+|\n\psi|^{\frac 4 3}].$$
Our aim is to extend the previous Liouville theorems to the case of Dirac-harmonic maps.
We will prove the following

\vskip12pt\noindent
{\bf Theorem 1.1.} {\it   Let $M^n$ be one of $\mathbb{R}^n$,  $\mathbb{H} ^n$, $\mathfrak{S}^n$, $n\geq 3$,
 $N$ be any Riemannian manifold. Let $\phi: M\to N$ be a map and $\psi\in\Gamma(\Sigma M\otimes\phi^{-1}TN)$.
 If $(\phi,\psi)$ is a Dirac-harmonic map with finite energy:
\begin{equation}
\label{b3} E(\phi,\psi):=\int[|d\phi|^2+|\psi|^4+|\n\psi|^{\frac 4
3}]<\infty,
\end{equation}
then $\phi$ must be constant and $\psi\equiv 0$.}
\vskip12pt

In fact, the supersymmetric $\sigma$-model in superstring theory
includes an additional curvature term in addition to (\ref{b0}).
Turning again  the components of $\psi$, which in quantum field
theory take values in some Grassmann algebra and anti-commute with
each other, into ordinary spinor fields on $M$, we have  the
following functional:
\begin{equation}
\label{b4} L_c(\phi,\psi):=\frac 1 2
\int_M[|d\phi|^2+\la\psi,\D\psi\ra-\frac 1 6 R_{ikjl}\la
\psi^i,\psi^j \ra \la \psi^k,\psi^l \ra].
\end{equation}
We call the critical points $(\phi,\psi)$ of $L_c$  {\it
Dirac-harmonic maps with curvature term}. We should point out that
the factor $-\frac 1 6$ in front of the curvature term in (\ref{b4})
is dictated by supersymmetry. Since in our treatment of the
functional, we shall not utilize this symmetry, the value of this
coupling constant will not be essential for us, except that changing
it from negative to positive values would also change the sign in
the curvature condition in Theorem 1.2
 below. In other words, with a positive instead of
a negative coupling constant, we would obtain a vanishing for
negatively curved targets.

The Euler-Lagrange equations of the functional $L_c$ are  (see section 2 below):
\begin{equation}
\label{3.2.11}
\tau^m(\phi)-\frac{1}{2}R^m\hskip0.000001mm_{lij}\la\psi^i,\nabla\phi^l\cdot\psi^j\ra+\frac{1}{12}h^{mp}R_{ikjl;p}
\la\psi^i,\psi^j\ra\la\psi^k,\psi^l\ra=0,
\end{equation}
\begin{equation}
\label{3.2.1}
\D\psi^m=\frac{1}{3}R^m\hskip0.000001mm_{jkl}\la\psi^j,\psi^l\ra\psi^k,
\qquad m=1,2,\cdots,n'.
\end{equation}

 For solutions of this system,
we also have a Liouville theorem. However, due to the presence of
the curvature term in the functional $L_c$, we will need a condition
on the curvature of the target $N$, namely that $N$ has
positive sectional curvature.

\vskip12pt\noindent
{\bf Theorem 1.2.}
{\it   Let $M$,
 $N$, $\phi$ and $\psi$
be as in Theorem 1.1, suppose $N$ has positive sectional curvature.
If $(\phi,\psi)$ is a Dirac-harmonic map with curvature term with
finite energy, then $\phi$ must be constant and $\psi\equiv 0$.}
\vskip48pt

\section{The Euler-Lagrange equations for $L_c$}
\addtocounter{equation}{-10}

Let us first derive the Euler-Lagrange equations for $L_c$. We put
$$A:= h_{ij}(\phi)g^{\alpha\beta}\frac{\partial
\phi^i}{\partial x_\alpha}\frac{\partial \phi^j}{\partial x_\beta},
\quad B:=h_{ij}(\phi)\la\psi^i,\D\psi^j\ra, \quad
R:=-\frac{1}{6}R_{ikjl}(\phi) \la\psi^i,\psi^j\ra
\la\psi^k,\psi^l\ra,  $$ and have
$$
L_c = \frac{1}{2}\int_M (A+B+R).
$$
First, noting that
 \begin{eqnarray*}
 \delta_\psi B&=&2\la\delta\psi,\D\psi\ra\\
 &=&2h_{ij}\la\delta\psi^i,\D\psi^j\ra
 \end{eqnarray*}
and
\begin{eqnarray*}
\delta_\psi R &=&
-\frac{1}{6}R_{ijkl}[\la\delta\psi^i,\psi^k\ra\la\psi^j,\psi^l\ra+\la\psi^i,\delta\psi^k\ra \la\psi^j,\psi^l\ra\\
&&
+\la\psi^i,\psi^k\ra\la\delta\psi^j,\psi^l\ra+\la\psi^i,\psi^k\ra \la\psi^j,\delta\psi^l\ra]\\
&=&-\frac{2}{3}R_{ijkl}\la\delta\psi^i,\psi^k\ra
\la\psi^j,\psi^l\ra,
\end{eqnarray*}
we have
\begin{eqnarray*}
\delta_\psi L_c &=&
\frac{1}{2}\int_M[2h_{ij}\la\delta\psi^i,\D\psi^j\ra-\frac{2}{3}R_{ijkl}\la\delta\psi^i,\psi^k\ra
\la\psi^j,\psi^l\ra]
\\
&=& \int_M [\la\delta
\psi^i,h_{ij}\D\psi^j\ra-\frac{1}{3}R_{ijkl}\la\delta\psi^i,\psi^k\ra
\la\psi^j,\psi^l\ra],
\end{eqnarray*}
which implies that
$$h_{ij}\D\psi^j-\frac{1}{3}R_{ijkl}\psi^k\la\psi^j,\psi^l\ra=0.$$
Thus, we obtain {\bf \it the $\psi$-equation for $L$}:
\begin{equation}
\label{3.2.1}
\D\psi^m=\frac{1}{3}R^m\hskip0.000001mm_{jkl}\la\psi^j,\psi^l\ra\psi^k.
\end{equation}
Second, consider the $\phi$-variation $\{\phi_t\}$ with
$\phi_0=\phi$ and $\frac{d\phi_t}{dt}|_{t=0}=\xi,$ we have
\begin{eqnarray}
\label{3.2.2}
\frac{dL_c(\phi_t)}{dt}|_{t=0}&=&\frac{1}{2}\int_M\frac{\partial}{\partial
t}|d\phi_t|^2|_{t=0} +\frac{1}{2}\int_M\frac{\partial}{\partial
t}\la\psi,\D\psi\ra|_{t=0} \nonumber
\\
&&-\frac{1}{12}\int_M\frac{\partial}{\partial
t}(R_{ijkl}\la\psi^i,\psi^k\ra\la\psi^j,\psi^l\ra)|_{t=0} \nonumber \\
&:=& I_1+I_2+I_3.
\end{eqnarray}
For the term $I_1$ it is well-known that (see e.g. \cite{Xin} or
\cite{J})
\begin{equation}
\label{3.2.3} I_1=-\int_M h_{im}\tau^i(\phi)\xi^m.
\end{equation}
For $I_2$ we choose an orthonormal basis
$\{e_\alpha|\alpha=1,2,\cdots,n\}$ on $M$ with
$[e_\alpha,\partial_t]=0$. Note that
\begin{eqnarray}
\label{3.2.4} \frac{\partial}{\partial t}\la\psi,\D\psi\ra&=&
\la\widetilde{\nabla}_{\partial_t}\psi,\D\psi\ra+\la\psi,\widetilde{\nabla}_{\partial_t}\D\psi\ra
\nonumber \\
&:=&
\la\psi_t,\D\psi\ra+\la\psi,\widetilde{\nabla}_{\partial_t}\D\psi\ra.
\end{eqnarray}
One can compute
\begin{eqnarray*}
\widetilde{\nabla}_{\partial_t}\D\psi&=&
\widetilde{\nabla}_{\partial_t}(e_\alpha\cdot\wa\psi) \\
&=& e_\alpha\cdot
\nabla_{e_\alpha}\psi^i\otimes\nabla_{\partial_t}\partial_{
y^i}+e_\alpha\cdot\psi^i\otimes\nabla_{\partial_t}\nabla_{e_\alpha}\partial_{
y^i}\\
&=&
e_\alpha\cdot\nabla_{e_\alpha}\psi^i\otimes\nabla_{\partial_t}\partial_{
y^i}+e_\alpha\cdot\psi^i\otimes[\nabla_{e_\alpha}\nabla_{\partial_t}\partial_{y^i}+R(\partial_t,e_\alpha)\partial_{y^i}]\\
&=& e_\alpha\cdot\wa(\psi^i\otimes\nabla_{\partial_t}\partial_{
y^i})+e_\alpha\cdot\psi^i\otimes
R^N(d\phi(\partial_t),d\phi(e_\alpha))\partial_{y^i} \\
&=&\D\psi_t+e_\alpha\cdot\psi^i\otimes
R^N(d\phi(\partial_t),d\phi(e_\alpha))\partial_{y^i}.
\end{eqnarray*}
It follows that
\begin{equation}
\label{3.2.5}
\la\psi,\widetilde{\nabla}_{\partial_t}\D\psi\ra=\la\psi,\D\psi_t\ra+\la\psi,e_\alpha\cdot\psi^i\otimes
R^N(d\phi(\partial_t),d\phi(e_\alpha))\partial_{y^i}\ra.
\end{equation}
Since
\begin{eqnarray*}
R^N(d\phi(\partial_t),d\phi(e_\alpha))\partial_{y^i}|_{t=0}&=&
R^N(\xi^m\partial_{y^m},\phi^l_\alpha\partial_{y^l})\partial_{y^i}\\
&=& \xi^m\phi^l_\alpha R^j_{iml}\partial_{y^j},
\end{eqnarray*}
we have
\begin{eqnarray*}
\la\psi,e_\alpha\cdot\psi^i\otimes
R^N(d\phi(\partial_t),d\phi(e_\alpha))\partial_{y^i}\ra|_{t=0}&=&
\la\psi,\xi^m\phi^l_\alpha R^j_{iml}\partial_{y^j}\otimes
e_\alpha\cdot\psi^i\ra\\
&=& \la\psi^i,\nabla\phi^l\cdot\psi^j\ra R_{mlij}\xi^m.
\end{eqnarray*}
From this formula and (\ref{3.2.5}) we have
$$\la\psi,\widetilde{\nabla}_{\partial_t}\D\psi\ra|_{t=0}=\la\psi,\D\psi_t\ra|_{t=0}+\la\psi^i,\nabla\phi^l\cdot\psi^j\ra R_{mlij}\xi^m.$$
Combining this with (\ref{3.2.4}) we obtain
$$\frac{\partial}{\partial t}\la\psi,\D\psi\ra|_{t=0}=
\la\psi_t,\D\psi\ra|_{t=0}+\la\psi,\D\psi_t\ra|_{t=0}+\la\psi^i,\nabla\phi^l\cdot\psi^j\ra
R_{mlij}\xi^m.$$ Thus, we have
\begin{equation}
\label{3.2.6}
I_2=\frac{1}{2}\int_M[\la\psi_t,\D\psi\ra+\la\psi,\D\psi_t\ra]|_{t=0}+\frac{1}{2}\int_M\la\psi^i,\nabla\phi^l\cdot\psi^j\ra
R_{mlij}\xi^m.
\end{equation}
From
$$\psi_t=\widetilde{\nabla}_{\partial_t}(\psi^i\otimes\partial_{y^i})|_{t=0}=
\psi^i\otimes\nabla_{d\phi(\partial_t)}\partial_{y^i}|_{t=0}=\xi^m\psi^i\otimes\Gamma^k_{im}\partial_{y^k},$$
we have
\begin{eqnarray*}
\la\psi_t,\D\psi\ra|_{t=0}&=&\la\xi^m\psi^i\Gamma^k_{im}\otimes\partial_{y^k},\D\psi^l\otimes\partial_{y^l}\ra
\\
&=&\la\xi^m\psi^i\Gamma^k_{im},\D\psi^lh_{kl}\ra\\
&=&\la\psi^i,\D\psi^j\ra\xi^m\Gamma_{im,j},
\end{eqnarray*}
where $\Gamma_{im,j}:=\Gamma^k_{im}h_{kj}$. Therefore,
\begin{equation}
\label{3.2.7}
I_2=\int_M\la\psi^i,\D\psi^j\ra\xi^m\Gamma_{im,j}+\frac{1}{2}\int_M\la\psi^i,\nabla\phi^l\cdot\psi^j\ra
R_{mlij}\xi^m.
\end{equation}
From (\ref{3.2.3}) and (\ref{3.2.7}) we obtain
\begin{equation}
\label{3.2.8} I_1+I_2=\frac{1}{2}\int_M[-2 h_{im}\tau^i(\phi)+2
 \la\psi^i,\D\psi^j\ra\xi^m\Gamma_{im,j}+\la\psi^i,\nabla\phi^l\cdot\psi^j\ra R_{mlij}]\xi^m.
\end{equation}
Using the $\psi$-equation,
\begin{eqnarray*}
2 \la\psi^i,\D\psi^j\ra\xi^m\Gamma_{im,j}&=&
   2\la\psi^i,\D\psi^p\ra\xi^m\Gamma_{im,p}\\
   &=& \frac{2}{3}\Gamma_{mi,p}R^p_{jkl}\la\psi^i,\psi^k\ra\la\psi^j,\psi^l\ra,
   \end{eqnarray*}
we have
\begin{equation}
\label{3.2.9} I_1+I_2=\frac{1}{2}\int_M[-2
h_{im}\tau^i(\phi)+\frac{2}{3}\Gamma_{mi,p}R^p\hskip0.000001mm_{jkl}\la\psi^i,\psi^k\ra
\la\psi^j,\psi^l\ra +\la\psi^i,\nabla\phi^l\cdot\psi^j\ra
R_{mlij}]\xi^m.
\end{equation}
The term $I_3$ is easy to compute.
$$I_3=-\frac{1}{2}\int_M\frac{1}{6}R_{ijkl,m}\la\psi^i,\psi^k\ra \la\psi^j,\psi^l\ra\xi^m.$$
Substituting this and (\ref{3.2.9}) into (\ref{3.2.2}) yields
\begin{eqnarray}
\label{3.2.10}
 \frac{dL_c(\phi_t)}{dt}|_{t=0}&=& \frac{1}{2}\int_M[-2
h_{im}\tau^i(\phi)+\frac{2}{3}\Gamma_{mi,p}R^p\hskip0.000001mm_{jkl}\la\psi^i,\psi^k\ra\la\psi^j,\psi^l\ra
\nonumber \\
&&+\la\psi^i,\nabla\phi^l\cdot\psi^j\ra
R_{mlij}-\frac{1}{6}R_{ijkl,m}\la\psi^i,\psi^k\ra\la\psi^j,\psi^l\ra]
\xi^m\nonumber\\
&=& \frac{1}{2}\int_M[-2
h_{im}\tau^i(\phi)+\la\psi^i,\nabla\phi^l\cdot\psi^j\ra R_{mlij}\nonumber\\
&&-\frac{1}{6}R_{ijkl;m}\la\psi^i,\psi^k\ra\la\psi^j,\psi^l\ra]\xi^m.
\end{eqnarray}
Here, $R_{ijkl;m}$ denotes the covariant derivative of the curvature
tensor $R_{ijkl}$ with respect to $\frac{\partial}{\partial y^m}$.
Therefore, we obtain {\bf \it the $\phi$-equation for $L_c$}:
\begin{equation}
\label{3.2.11}
\tau^m(\phi)-\frac{1}{2}R^m\hskip0.000001mm_{lij}\la\psi^i,\nabla\phi^l\cdot\psi^j\ra+\frac{1}{12}h^{mp}R_{ikjl;p}
\la\psi^i,\psi^j\ra\la\psi^k,\psi^l\ra=0.
\end{equation}

\vskip48pt

\section{Proofs of theorems}
\addtocounter{equation}{-11}

Now we start to prove our main Theorems. Suppose $X\in \Gamma(TM)$
is a conformal vector field on $(M,g)$, namely,
\begin{equation}
\label{b2.1}
L_X g=2fg,
\end{equation}
where $f\in C^\infty (M)$. Here $L_X$ denotes the Lie derivative
with respect to $X$. The vector field $X$ generates a family of
conformal diffeomorphisms
$$F_t:=exp(tX): M\to M.$$
We will consider the variation of the functionals $L$ and $L_c$
under this family of diffeomorphisms.

\par
In the Euclidean space $\R^n$, the vector field $X(x):=x$ is
conformal with  $f=1$. Consider $\R^n$ equipped with a metric
$$g=b^2(dr^2+a^2 d\Theta ^2),$$
where $a,b$ are radial functions, $(r,\Theta )$ are polar
coordinates centered at the origin, and $d\Theta ^2$ stands for
the standard metric on the unit sphere $\S^{n-1}$. Then the vector
field $X:=a(r)\partial_r$ satisfies:  $L_X g=2fg$ with
$f=(ab)'/b$, that is, $X$ is a conformal vector field
 (c.f. \cite{Pk}).
Besides the standard Euclidean space $\R^n$, we also consider the
following cases:
 \vskip12pt
\par
\noindent (i) The hyperbolic space $\H^n=\{(x,t)\in \R^n\times
\R|1+x^2=t^2\}$:  $b^2=1/(r^2+1)$ and

$a=r/b$.  In this case, $f\geq 1$ and $|X(r)|\leq r$.
\par
\vskip12pt \noindent (ii)  $\mathfrak{S}^n$ with the Schwarzschild
metric: a constant slice of the outer region $(r>r_0$

$:=2m)$ of $n+1$-dimensional Schwarzschild space,
$b=1/\sqrt{1-\frac{r_0}{r}}$ and $a=r/b$,

where $m$ is the mass of a black hole. In this case, $0<f\leq 1$
and $|X(r)|\leq r.$

\vskip12pt Recall the definition of $L$:
$$L(\phi,\psi,g)=\frac 1 2\int_M[|d\phi|^2+\la\psi,\D\psi\ra]v_g,$$
where $v_g:=\sqrt{detg_{\alpha\beta)}}dx$ is the volume form of $M$.
\par
$$\Omega:=(|d\phi|^2+\la\psi,\D\psi\ra)v_g$$
is an $n$-form on $M$. We note that for any $\eta\in C_0^\infty (M),$
$$0=\int_Md[(\imath_X\Omega)\eta]=\int_M\eta d(\imath_X\Omega) +\int_Md\eta \wedge \imath_X\Omega
=\int_M\eta L_X\Omega +\int_Md\eta \wedge \imath_X\Omega,$$
that is,
\begin{equation}
\label{b2.2} \int_M\eta L_X\Omega=-\int_Md\eta \wedge
\imath_X\Omega,
\end{equation}
where $\imath_X$ stands for the inner product with the vector
$X$.

Now let us compute $L_X\Omega$. We first recall the following
\vskip12pt \noindent {\bf Lemma 2.1(c.f. \cite{EL}).} {\it Let
$\phi: M\to N$ be a map, and $X$ any smooth vector field on $M$.
Then
\begin{equation}
\label{b2.3}
L_X(\frac 1 2|d\phi|^2 v_g)=\la d\phi,\nabla(d\phi(X))\ra v_g+\frac 1 2 \la L_Xg,S_\phi\ra v_g,
\end{equation}
\begin{equation}
\label{b2.4}
L_Xv_g=\frac 1 2 \la L_X g,g\ra v_g,
\end{equation}
where $S_\phi:=\frac 1 2 |d\phi|^2g-\phi^*h$ is the stress-energy
tensor of $\phi$. } \vskip12pt
 Second, we note that
\begin{eqnarray}
\label{b2.5}
L_X(\la \D\psi,\psi\ra v_g)&=&(L_X\la \D\psi,\psi\ra )v_g+\la \D\psi,\psi\ra L_Xv_g\nonumber \\
&=&\la L_X(\D\psi),\psi\ra v_g+\la \D\psi,L_X\psi\ra v_g+\frac 1 2
\la \D\psi,\psi\ra \la L_X g,g\ra v_g.
\end{eqnarray}
To continue, we recall that
\begin{eqnarray*}
\D\psi&=&e_\alpha\cdot\widetilde{\n}_{e_\alpha}\psi\\
&=&\p
\psi^i\otimes\partial_{y^i}(\phi)+(e_\alpha\cdot\psi^i)\phi^j_\alpha\n_{\partial_{y^j}}\partial_{y^i}(\phi),
\end{eqnarray*}
The variation of $\D\psi$ consists of two parts: one with respect
to the metric $g$, the other with respect to the parameterization
$p$ of $M$ caused by $X$, namely,
\begin{equation}
\label{b2.5}
\frac{d}{dt}(\D\psi)|_{t=0}=\delta_g(\D\psi)+\delta_p(\D\psi).
\end{equation}
\vskip3mm
 \noindent {\bf Lemma 2.2.} {\it The first variation is:
\begin{equation}
\label{b2.6.1}\delta_g(\D\psi)=-\frac 1
2e_\alpha\cdot\widetilde{\n}_{K(e_\alpha)}\psi+A\cdot
\psi,\end{equation}
 where $A:=\frac 1 4 [{\rm div}_g k+d({\rm
Tr}_g k)]$, $k:=L_Xg$ and $K$ is a (1,1)-tensor on $M$ defined by
$$g(K(e_\alpha),e_\beta):=k(e_\alpha,e_\beta)=L_Xg(e_\alpha,e_\beta).$$

} \vskip12pt
 \par \noindent {\it Proof.} The proof follows closely \cite{BG}. See also
 \cite{Maier}. In order
to obtain (\ref{b2.6.1}), we first note that given any real
$n-$dimensional vector space $V$ equipped with a metric $g$, then
for any other metric $g'$ on $V$, there exists a unique positive
endomorphism $H$ on $V$ such that
$g'(\cdot,\cdot)=g(H(\cdot),\cdot)$. It
is clear that $b_{g',g}:=H^{-1/2}$ transforms $g-$orthonormal frames to
$g'-$orthonormal frames. And consequently, we have an
$SO_n$-equivariant map from the manifold $P(g)$ of $g-$orthonormal
frames
 to the manifold  $P(g')$ of $g'-$orthonormal frames.

 Since $M$ is spin, the map $b_{g',g}$ can
be lifted to a $Spin_n$-equivariant map $\beta_{g',g}:
\tilde{P}(g)\to \tilde{P}(g')$. Extend $b_{g',g}$ and $\beta_{g',g}$
to an
 $SO_n$-equivariant map $b_{g',g}:P_{SO}(M,g)\to
P_{SO}(M,g')$ and a $Spin_n$-equivariant map
 $\beta_{g',g}:P_{Spin}(M,g)\to P_{Spin}(M,g')$ respectively.
 Denote the spin bundles with respect to $g$ and $g'$ by $\Sigma_gM$ and $\Sigma_{g'}M$ respectively, then the map
$\beta_{g',g}$ extends to an isometry $\beta_{g',g}: \Sigma_gM\to
\Sigma_{g'}M$ of Hermitian bundles. Clearly,
$\beta^{-1}_{g',g}=\beta_{g,g'}.$

For the Dirac operator $\D$, we consider the transformation operator
acting on the spin bundle $\Sigma_gM$:
$$\D\hskip0.5mm_{g',g}:=\beta^{-1}_{g',g}\D\hskip0.5mm_{g'}\beta_{g',g},$$
where $\D\hskip0.5mm_{g'}$ denotes the Dirac operator $\D$ with
respect to the metric $g'$ on $M$, namely,
\begin{eqnarray*}
\D\hskip0.5mm_{g'}\psi &=&\p\hskip0.5mm_{g'}
\psi^i\otimes\partial_{y^i}(\phi)+(e_{\alpha,g'}\cdot\psi^i)\otimes\n_{e_{\alpha,g'}}\phi^j\n_{\partial_{y^j}}\partial_{y^i}(\phi)\\
&=&\p\hskip0.5mm_{g'} \psi^i\otimes\partial_{y^i}(\phi)+({\rm
grad}_{g'}\phi^j\cdot\psi^i)\otimes\n_{\partial_{y^j}}\partial_{y^i}(\phi);
\end{eqnarray*}
here, $\{e_{\alpha,g'}\}$ denotes the $g'-$orthornormal frame,
which is transformed via $b_{g,g'}$ to the $g-$orthornormal frame
 $\{e_{\alpha}\}$:
 $$b_{g,g'}(e_{\alpha,g'})=e_{\alpha}.$$
 Hence,
\begin{eqnarray}
\label{b2.6.3}
\D\hskip0.5mm_{g',g}\psi&=&\beta^{-1}_{g',g}\p\hskip0.5mm_g'\beta_{g',g}\psi^i\otimes\partial_{y^i}+\beta^{-1}_{g',g}({\rm
grad}_{g'}\phi^j)\beta_{g',g}\cdot\psi^i\otimes\n_{\partial_{y^j}}\partial_{y^i}\nonumber\\
&=&\p_{g',g}\psi^i\otimes\partial_{y^i}+b_{g,g'}({\rm
grad}_{g'}\phi^j)\cdot\psi^i\otimes\n_{\partial_{y^j}}\partial_{y^i}\nonumber\\
&=&\p_{g',g}\psi^i\otimes\partial_{y^i}+b_{g,g'}(\n_{e_{\alpha,g'}}
\phi^j
e_{\alpha,g'})\cdot\psi^i\otimes\n_{\partial_{y^j}}\partial_{y^i}\nonumber\\
&=&\p_{g',g}\psi^i\otimes\partial_{y^i}+\n_{e_{\alpha,g'}} \phi^j
(e_\alpha\cdot\psi^i)\otimes\n_{\partial_{y^j}}\partial_{y^i}.
\end{eqnarray}
For the variation $\{g_t\}$ of $g$ with
$\frac{dg_t}{dt}|_{t=0}=L_Xg$, we have
$$\D\hskip0.5mm_{g_t,g}\psi=\p_{g_t,g}\psi^i\otimes\partial_{y^i}+\n_{e_{\alpha,g_t}} \phi^j
(e_\alpha\cdot\psi^i)\otimes\n_{\partial_{y^j}}\partial_{y^i},$$
from which we have
\begin{equation}
\label{b2.6.4}
\frac{d}{dt}(\D\hskip0.5mm_{g_t,g}\psi)|_{t=0}=\frac{d}{dt}(\p_{g_t,g})|_{t=0}\psi^i\otimes\partial_{y^i}+\n_{\frac{d}{dt}(b_{g_t,g})|_{t=0}(e_{\alpha})}
\phi^j
(e_\alpha\cdot\psi^i)\otimes\n_{\partial_{y^j}}\partial_{y^i}.
\end{equation}
Since $b_{g_t,g}=(Id+tK)^{-1/2}$, it follows that
\begin{equation}
\label{b2.6.5} \frac{d}{dt}(b_{g_t,g})|_{t=0}=-\frac 1 2 K.
\end{equation}
On the other hand, Theorem 21 in \cite{BG} gives us
\begin{equation}
\label{b2.6.6} \frac{d}{dt}(\p_{g_t,g})|_{t=0}\psi^i=-\frac 1
2e_\alpha\cdot\n_{K(e_\alpha)}\psi^i+A\cdot\psi^i,\qquad
i=1,2,\cdots,n'.
\end{equation}
Inserting (\ref{b2.6.5}) and  (\ref{b2.6.6}) into (\ref{b2.6.4})
then yields
\begin{eqnarray*}
\frac{d}{dt}(\D\hskip0.5mm_{g_t,g}\psi)|_{t=0}&=&[-\frac 1
2e_\alpha\cdot\n_{K(e_\alpha)}\psi^i+A\cdot\psi^i]\otimes\partial_{y^i}-\frac
1 2 \n_{K(e_\alpha)}\phi^j
(e_\alpha\cdot\psi^i)\otimes\n_{\partial_{y^j}}\partial_{y^i}\\
&=&-\frac 1 2 e_\alpha\cdot[\n_{K(e_\alpha)}\psi^i\otimes\partial_{y^i}+\psi^i\otimes\n_{K(e_\alpha)}\phi^j
     \n_{\partial_{y^j}}\partial_{y^i}]+A\cdot\psi\\
&=&-\frac 1 2e_\alpha\cdot\widetilde{\n}_{K(e_\alpha)}\psi+A\cdot
\psi.
\end{eqnarray*}
This proves Lemma 2.2.                   \hfill Q.E.D. \vskip12pt
 Thus, from (\ref{b2.6.1}) we have
\begin{equation}
\label{b2.6} \delta_g(\D\psi)=-\frac 1
2(L_Xg)(e_\alpha,e_\beta)(e_\alpha\cdot\wb\psi)+A\cdot\psi.
\end{equation}
Now we compute the second variation
\begin{eqnarray}
\label{2.7} \delta_p
(\D\psi)&=&\delta_p[\p\psi^i\otimes\partial_{y^i}(\phi)+(e_\alpha\cdot\psi^i)\otimes
\n_{e_\alpha}\partial_{y^i}]\nonumber\\
&=&\p\psi^i\otimes\delta_\phi(\partial_{y^i}(\phi))+(e_\alpha\cdot\psi^i)\otimes\delta_\phi
(\n_{e_\alpha}\partial_{y^i})+\D(L_X\psi^i\otimes\partial_{y^i})\nonumber\\
&=&[\p\psi^i\otimes\n_{\partial_t}\partial_{y^i}+(e_\alpha\cdot\psi^i)\otimes\n_{\partial_t}
\n_{e_\alpha}\partial_{y^i}]|_{t=0}+\D(L_X\psi^i\otimes\partial_{y^i})\nonumber\\
&=&[\p\psi^i\otimes\n_{\partial_t}\partial_{y^i}+(e_\alpha\cdot\psi^i)\otimes\n_{e_\alpha}
(\n_{\partial_t}\partial_{y^i})\nonumber\\
&&+(e_\alpha\cdot\psi^i)\otimes
R^N(\frac{d\phi}{dt},\phi_\alpha)\partial_{y^i}]|_{t=0}+\D(L_X\psi^i\otimes\partial_{y^i})\nonumber\\
&=&\D(L_X\psi)+[(e_\alpha\cdot\psi^i)\otimes
R^N(\frac{d\phi}{dt},\phi_\alpha)\partial_{y^i}]|_{t=0}.
\end{eqnarray}
Therefore,
\begin{eqnarray}
\label{2.8}
L_X(\D\psi)&=&\delta_g(\D\psi)+\delta_p(\D\psi)\nonumber\\
&=&-\frac 1
2(L_Xg)(e_\alpha,e_\beta)(e_\alpha\cdot\wb\psi)+A\cdot\psi\nonumber\\
&&+\D(L_X\psi)+[(e_\alpha\cdot\psi^i)\otimes
R^N(\frac{d\phi}{dt},\phi_\alpha)\partial_{y^i}]|_{t=0}\nonumber\\
&=&-\frac 1
2(L_Xg)(e_\alpha,e_\beta)(e_\alpha\cdot\wb\psi)+A\cdot\psi\nonumber\\
&&+\D(L_X\psi)+(e_\alpha\cdot\psi^i)\otimes
R^N(d\phi(X),d\phi(e_\alpha))\partial_{y^i},
\end{eqnarray}
from which we have
\begin{eqnarray}
\label{b2.9} \la L_X(\D\psi),\psi\ra&=&-\frac 1 2
(L_Xg)(e_\alpha,e_\beta)\la
e_\alpha\cdot\wb\psi,\psi\ra+\la\D(L_X\psi),\psi\ra\nonumber\\
&&+\la(e_\alpha\cdot\psi^i)\otimes
R^N(d\phi(X),d\phi(e_\alpha))\partial_{y^i},\psi\ra.
\end{eqnarray}
The last term in the above equality can be calculated as follows
\begin{eqnarray}
\label{b2.10} \la(e_\alpha\cdot\psi^i)\otimes
R^N(d\phi(X),d\phi(e_\alpha))\partial_{y^i},\psi\ra&=&\la
e_\alpha\cdot\psi^i,\psi^j\ra\la
R^N(\partial_{y^m},\partial_{y^l})\partial_{y^i},\partial_{y^j}\ra
X(\phi^m)\phi^l_\alpha\nonumber\\
&=&\la\n\phi^l\cdot\psi^j,\psi^i\ra R_{mlij} X(\phi^m)\nonumber\\
&=&2\la \frac 1 2 R^k\hskip0.0005mm
_{lij}\la\psi^i,\n\phi^l\cdot\psi^j\ra\partial_{y^k},d\phi(X)\ra\nonumber\\
&=&2\la \mathcal{R}(\phi,\psi),d\phi(X)\ra.
\end{eqnarray}
Thus, we have
\begin{eqnarray}
\label{b2.11} \la L_X(\D\psi),\psi\ra&=&-\frac 1 2
(L_Xg)(e_\alpha,e_\beta)\la
e_\alpha\cdot\wb\psi,\psi\ra+\la\D(L_X\psi),\psi\ra\nonumber\\
&&+2\la \mathcal{R}(\phi,\psi),d\phi(X)\ra.
\end{eqnarray}
Finally, we have
\begin{eqnarray}
\label{b2.12} L_X(\la \D\psi,\psi\ra v_g)&=&-\frac 1 2
(L_Xg)(e_\alpha,e_\beta)\la e_\alpha\cdot\wb\psi,\psi\ra
v_g+\la\D(L_X\psi),\psi\ra v_g\nonumber\\
&&+2\la \mathcal{R}(\phi,\psi),d\phi(X)\ra v_g+\la
\D\psi,L_X\psi\ra v_g+\frac 1 2 \la\D\psi, \psi\ra\la L_Xg,g\ra
v_g.
\end{eqnarray}

\vskip24pt

\noindent {\it Proof of Theorem 1.1.} Assume that $X$ is a conformal
vector field: $L_X g=2fg,$ and $(\phi,\psi)$ is a Dirac-harmonic
map: $\tau (\phi)=\cal{R}(\phi,\psi),$ $\D\psi=0.$ Then from
(\ref{b2.12}) we have
\begin{eqnarray}
\label{b2.13} L_X(\la \D\psi,\psi\ra v_g)&=&-f\la\D\psi, \psi\ra
v_g+\la\D(L_X\psi),\psi\ra v_g\nonumber\\
&&+2\la \mathcal{R}(\phi,\psi),d\phi(X)\ra v_g\nonumber\\
&=&\la\D(L_X\psi),\psi\ra v_g+2\la
\mathcal{R}(\phi,\psi),d\phi(X)\ra v_g.
\end{eqnarray}
From Lemma 2.1, we have
\begin{eqnarray}
\label{b2.14} L_X(|d\phi|^2 v_g)&=&2\la d\phi,\nabla(d\phi(X))\ra
v_g+f \la g,S_\phi\ra v_g\nonumber\\
&=&2\la d\phi,\nabla(d\phi(X))\ra v_g+ \frac{n-2}{2} f|d\phi|^2
v_g.
\end{eqnarray}
Combining (\ref{b2.13}) and (\ref{b2.14}) yields
\begin{eqnarray}
\label{b2.15} \int_M\eta L_X \Omega
&=&\int_M\eta\la\D(L_X\psi),\psi\ra v_g+\frac{n-2}{2}\int_M\eta
f|d\phi|^2
v_g\nonumber\\
&&+2 \int_M\eta \la d\phi,\nabla(d\phi(X))\ra v_g+2 \int_M\eta\la
\mathcal{R}(\phi,\psi),d\phi(X)\ra v_g.
\end{eqnarray}
Note that
\begin{eqnarray}
\label{b2.16} \int_M\eta \la d\phi,\nabla(d\phi(X))\ra
v_g&=&\int_M\eta \la d\phi(e_\alpha),\nabla e_\alpha(d\phi(X))\ra
v_g\nonumber\\
&=&\int_M\n_{e_\alpha}(\eta\la d\phi(e_\alpha),d\phi(X)\ra)v_g
-\int_M \la d\phi(\n\eta),d\phi(X)\ra v_g\nonumber\\
&&-\int \eta \la \n_{e_\alpha} d\phi(e_\alpha),d\phi(X)\ra
v_g\nonumber\\
 &=& -\int_M\la d\phi(\n\eta),d\phi (X)\ra v_g-\int_M \eta\la \tau
(\phi),d\phi(X)\ra v_g.
\end{eqnarray}
Putting this into (\ref{b2.15}), we obtain
\begin{eqnarray}
\label{b2.17} \int_M\eta L_X \Omega
&=&\int_M\eta\la\D(L_X\psi),\psi\ra v_g+\frac{n-2}{2}\int_M\eta
f|d\phi|^2
v_g\nonumber\\
&& -2\int_M\la d\phi(\n\eta),d\phi (X)\ra v_g-2\int \eta\la \tau
(\phi)-\cal{R}(\phi,\psi),d\phi(X)\ra v_g\nonumber\\
&=&\int_M\eta\la\D(L_X\psi),\psi\ra v_g+\frac{n-2}{2}\int_M\eta
f|d\phi|^2
v_g\nonumber\\
&& -2\int_M\la d\phi(\n\eta),d\phi (X)\ra v_g.
\end{eqnarray}
But
\begin{eqnarray}
\label{b2.18} \int_M\eta\la\D(L_X\psi),\psi\ra v_g&=&\int_M\la
L_X\psi,\D(\eta\psi)\ra v_g\nonumber\\
&=&\int_M\la L_X\psi,\n\eta\cdot\psi+\eta\D\psi\ra v_g\nonumber\\
&=&\int_M\la L_X\psi,\n\eta\cdot\psi\ra v_g,
\end{eqnarray}
therefore,
\begin{eqnarray}
\label{b2.19} \int_M\eta L_X \Omega &=&\int_M\la
L_X\psi,\n\eta\cdot\psi\ra v_g+\frac{n-2}{2}\int_M\eta f|d\phi|^2
v_g\nonumber\\
&& -2\int_M\la d\phi(\n\eta),d\phi (X)\ra v_g.
\end{eqnarray}
Using the equation (\ref{b1}), i.e., $\D\psi=0$, we have
\begin{equation}
\label{b2.20}
-\int_M d\eta\wedge \imath_X\Omega =-\int_M d\eta\wedge \imath_X(|d\phi|^2 v_g).
\end{equation}
Putting (\ref{b2.19}) and (\ref{b2.20}) into (\ref{b2.2}) yields
\begin{eqnarray}
\label{b2.21}
\frac{n-2}{2}\int_M\eta f |d\phi|^2&=&2\int_M\la d\phi(\n\eta,d\phi(X)\ra v_g-\int_M\la L_X\psi,
\n\eta\cdot\psi\ra v_g\nonumber\\
&&-\int_M(d\eta\wedge \imath_X v_g)|d\phi|^2.
\end{eqnarray}
\vskip12pt \noindent (1) $M=\R^n, \H^n (n\geq 3)$:   For any
$R>0$, choose a cut-off function $\eta_R$ such that $0\leq
\eta_R\leq 1$,
$$
            \eta_R=\left\{
                \begin{array}{l}
  1\quad B_R, \\
  0\quad M\setminus B_{2R},
                \end{array}
                \right.
  $$
and $|\eta_R'|\leq 2/R$. Inserting this into (\ref{b2.21}) yields
\begin{eqnarray}
\label{b2.22} \frac{n-2}{2}\int_M\eta_R f |d\phi|^2&\leq &
C[\int_{B_{2R}\setminus
B_R}(|d\phi|^2+|d\phi||\psi|^2+|\psi||\n\psi|)]\nonumber\\
&\leq& C\int_{B_{2R}\setminus
B_R}(|d\phi|^2+|\psi|^4+|\n\psi|^{\frac 4 3}).
\end{eqnarray}
Now in the previous formula letting $R\to+\infty$ and using the
finiteness of the energy, we have
$$\int_M f|d\phi|^2=0$$
which implies $\phi\equiv const.$ for $f>0$.

Now we fix coordinates $(y^i)$ at $\phi(M)$. Then from the
$\psi$-equation: $\D\psi=0$ we have
$$\p \psi^i=0,\qquad \int_M|\psi^i|^4 <\infty, \quad i=1,2,\cdots,
n'.$$ Denote $\xi:=\psi^i$, then $\xi\in \Gamma (\Sigma M)$ and
\begin{equation}
\label{b2.23} \p \xi=0,\qquad \int_M|\xi|^4 <\infty.
\end{equation}
By the Weitzenb\"{o}ck formula:
$$\frac 1 2 \Delta |\xi|^2=|\n\xi|^2+\frac 1 4 R_M |\xi|^2,$$
we have
$$\Delta |\xi|^2\geq -C|\xi|^2,$$
 where $R_M$ is the scalar curvature of $M$.

By a Morrey-type estimate (see e.g. \cite{Mo}, Theorem 5.3.1), we
conclude that for any $x_0\in M$ and $\rho>0$,
$$\sup\limits_{B_{x_0}(\rho)}|\xi|^4\leq\frac{C}{R^n}\int_{B_{x_0}(\rho+R)}|\xi|^4
\to 0 \quad (R\to +\infty),
$$
hence $\xi\equiv 0$ on $M$. We have proved the theorem for the
euclidean and hyperbolic case.
\par
\noindent (2) $M=\mathfrak{S}^n (n\geq 3, r>r_0):$  For any
$R>>1$, choose a cut-off function $\eta_R$ as in (1), and another
cut-off function $\zeta_\varepsilon$ such that
$$
\zeta_\varepsilon=\left\{
                \begin{array}{l}
  0\quad B_{r_0+\varepsilon}, \\
  1\quad M\setminus B_{r_0+2\varepsilon},
                \end{array}
                \right.
$$
and $|d\zeta_\varepsilon|\leq 2/\varepsilon.$ The functions $d\eta_R$
and $d\zeta_\varepsilon$ are supported in $B_{2R}\setminus B_R$ and
$B_{r_0+2\varepsilon}\setminus B_{r_0+\varepsilon}$ respectively.
Using $\eta=\eta_R\zeta_\varepsilon$ in (\ref{b2.21}), similar to
(\ref{b2.22}), we have
\begin{eqnarray}
\label{b2.24} \frac{n-2}{2}\int_{B_R\setminus
B_{r_0+2\varepsilon}} f |d\phi|^2&\leq &C[\int_{B_{2R}\setminus
B_R}(|d\phi|^2+|\psi|^4+|\n\psi|^{\frac 4 3})\nonumber\\
&&+\int_{B_{r_0+2\varepsilon}\setminus
B_{r_0+\varepsilon}}(|d\phi|^2+|\psi|^4+|\n\psi|^{\frac 4 3})].
\end{eqnarray}
Letting $R\to +\infty$ and $\varepsilon \to 0$, we obtain
$$\int_M f|d\phi|^2=0$$
which implies $\phi\equiv const.$ Similar to (1), we then conclude
that $\psi\equiv 0$ on $M$. This completes the proof of Theorem 1.1.
\hfill Q.E.D.

\vskip24pt \noindent {\it Proof of Theorem 1.2.}     Denote
$$\Omega_c:=(|d\phi|^2+\la\psi,\D\psi\ra-\frac 1 6 R_{ikjl}\la\psi^i,\psi^j\ra
\la\psi^k,\psi^l\ra)v_g.
$$
Similar to (\ref{b2.2}) we have
\begin{equation}
\label{b2.25} \int_M\eta L_X\Omega_c=-\int_Md\eta \wedge
\imath_X\Omega_c.
\end{equation}
As for the left hand side of this equality, from (\ref{b2.12}) and  the
conformality of the vector field $X$, we have
\begin{eqnarray}
\label{b2.26} L_X(\la \D\psi,\psi\ra v_g)&=&\la\D(L_X\psi),\psi\ra
v_g +2\la \mathcal{R}(\phi,\psi),d\phi(X)\ra v_g\nonumber\\
&&+\la \D\psi,L_X\psi\ra v_g+(n-1)f \la\D\psi, \psi\ra v_g.
\end{eqnarray}
Using (\ref{b2.4}) and the conformality of $X$ again, we have
\begin{eqnarray}
\label{b2.27} L_X(-\frac 1 6
R_{ikjl}\la\psi^i,\psi^j\ra\la\psi^k,\psi^l\ra v_g)&=&-\frac 1 6
R_{ikjl;p}X(\phi^p)\la\psi^i,\psi^j\ra\la\psi^k,\psi^l\ra
v_g\nonumber\\
&&-\frac 2 3R_{ikjl}\la L_X\psi^i,\psi^j\ra\la\psi^k,\psi^l\ra
v_g\nonumber\\
&&-\frac 2 3 \la \psi^i\otimes L_X\partial_{y^i},
R^m\hskip0.00005mm_{klj}\la\psi^k,\psi^l\ra\psi^j\partial_{y^m}\ra
\nonumber\\
&&- \frac 1 6 n f R_{ikjl}\la\psi^i,\psi^j\ra\la\psi^k,\psi^l\ra
v_g.
\end{eqnarray}
Recall (\ref{b2.14}):
\begin{eqnarray}
\label{b2.28} L_X(|d\phi|^2 v_g)&=&2\la d\phi,\nabla(d\phi(X))\ra
v_g+f \la g,S_\phi\ra v_g\nonumber\\
&=&2\la d\phi,\nabla(d\phi(X))\ra v_g+ \frac{n-2}{2} f |d\phi|^2
v_g.
\end{eqnarray}
Noting that
\begin{eqnarray}
\label{2.29} \int_M\la \D(L_X\psi),\eta\psi\ra v_g&=&\int_M \la
L_X\psi,\n\eta\cdot\psi+\eta\D\psi\ra v_g\nonumber\\
&=&\int_M \la L_X\psi,\n\eta\cdot\psi\ra v_g+\int_M \la
L_X\psi,\eta\D\psi\ra v_g,
\end{eqnarray}
and recall (\ref{b2.16}):
\begin{equation}
\label{b2.30} \int_M\eta \la d\phi,\nabla(d\phi(X))\ra
v_g=-\int_M\la d\phi(\n\eta),d\phi (X)\ra v_g-\int_M \eta\la \tau
(\phi),d\phi(X)\ra v_g,
\end{equation}
Combining (\ref{b2.26})-(\ref{b2.30}), and using the
Euler-Lagrange equations (\ref{3.2.1}) and (\ref{3.2.11}), we
obtain
\begin{eqnarray}
\label{b2.31} \int_M\eta L_X\Omega_c&=&\frac{n-2}{2}\int_M\eta f
|d\phi|^2 v_g +(n-1)\int_M\eta f \la \D\psi,\psi\ra
v_g\nonumber\\
&&-\frac n 6\int_M\eta f R_{ikjl}\la
\psi^i,\psi^j\ra\la\psi^k,\psi^l\ra v_g+\int_M\la L_X\psi,\n
\eta\cdot\psi\ra v_g\nonumber\\
&& -2\int_M\la d\phi(\n\eta),d\phi(X)\ra v_g.
\end{eqnarray}
On the other hand, the right hand side of (\ref{b2.25})
\begin{eqnarray}
\label{b2.32} -\int_Md\eta\wedge
\imath_X\Omega_c&=&-\int_M(d\eta\wedge \imath_X
v_g)(|d\phi|^2+\la\D\psi,\psi\ra-\frac 1 6 R_{ikjl}\la
\psi^i,\psi^j\ra\la\psi^k,\psi^l\ra)\nonumber\\
&=&-\int_M(d\eta\wedge \imath_X v_g)(|d\phi|^2+\frac 1 6 R_{ikjl}\la
\psi^i,\psi^j\ra\la\psi^k,\psi^l\ra).
\end{eqnarray}
Putting (\ref{b2.31}) and (\ref{b2.32}) into (\ref{b2.25}), and
using the same argument as in the proof of Theorem 1.1, we have
\begin{eqnarray}
\label{b2.33} \frac{n-2}{2}\int_M f |d\phi|^2 v_g
&+&(n-1)\int_M f \la \D\psi,\psi\ra v_g\nonumber\\
&& -\frac n 6\int_M f R_{ikjl}\la
\psi^i,\psi^j\ra\la\psi^k,\psi^l\ra v_g=0.
\end{eqnarray}
Substituting the $\psi$-equation (\ref{3.2.1}) into it,  we obtain
the following equality:
\begin{equation}
\label{b2.34} \int_M f |d\phi|^2 v_g +\frac 1 3\int_M f R_{ikjl}\la
\psi^i,\psi^j\ra\la\psi^k,\psi^l\ra v_g=0.
\end{equation}

Denote $a_{ij}:=\la\psi^i,\psi^j\ra$. The symmetric matrix
$(a_{ij})$ is semi-positive, therefore we can write
$$a_{ij}=b_{ip}b_{jp},$$
where $(b_{ij})$ is a real $n'\times n'$ matrix. Set
$b^p:=(b_{1p},b_{2p},\cdots,b_{n'p})$, then
\begin{eqnarray*}
R_{ikjl}\la
\psi^i,\psi^j\ra\la\psi^k,\psi^l\ra&=&R_{ikjl}a_{ij}a_{kl}\\
&=&R_{ikjl}b_{ip}b_{jp}b_{kq}b_{lq}\\
&=&R(b^p,b^q,b^p,b^q).
\end{eqnarray*}
Using the assumption that $N$ has positive sectional curvature and
noting that $f>0$, we immediately conclude that $\phi$ is constant
and $\psi$ vanishes. This completes the proof of Theorem 1.2. \hfill
Q.E.D.

\end{document}